\documentclass[aps,11pt,reprint,notitlepage,onecolumn,superscriptaddress,longbibliography]{revtex4-1}

\usepackage[english]{babel}
\usepackage[utf8]{inputenc}

\usepackage{amsmath}
\usepackage{amsthm}
\usepackage{amsfonts}
\usepackage{amssymb}
\usepackage{array}
\usepackage{bbold}
\usepackage{dsfont}
\usepackage{xcolor}
\definecolor{mypink}{RGB}{237, 152, 152}
\definecolor{myblue}{RGB}{161, 182, 242}
\usepackage{hyperref}
\hypersetup{colorlinks=true, citecolor=blue, linkcolor=blue,urlcolor= blue}
\usepackage{graphicx}
\usepackage{bm}

\newcommand{\ud}{\mathrm{d}}
\newcommand{\e}{\mathrm{e}}

\newcommand{\g}{\boldsymbol{\gamma}}

\newcommand{\R}{\mathds{R}}
\newcommand{\E}{\mathds{E}}
\newcommand{\A}{\boldsymbol{A}}
\newcommand{\C}{\boldsymbol{C}^n}
\newcommand{\bpi}{\boldsymbol{\pi}}
\newcommand{\ba}{\boldsymbol{a}}
\newcommand{\p}{\boldsymbol{p}}
\newcommand{\bmu}{\boldsymbol{\mu}}
\newcommand{\om}{\boldsymbol{\omega}}
\newcommand{\kk}{\boldsymbol{k}}
\newcommand{\bkappa}{\boldsymbol{\kappa}}
\newcommand{\K}{\boldsymbol{K}}
\newcommand{\uu}{\boldsymbol{u}}

\newcommand{\ff}{\boldsymbol{r}}

\newcommand{\h}{\boldsymbol{l}}

\newcommand{\M}{\boldsymbol{M}}
\newcommand{\N}{\boldsymbol{N}}

\newcommand{\bv}{\boldsymbol{v}}

\newcommand{\X}{\boldsymbol{X}}
\newcommand{\Y}{\boldsymbol{Y}}
\newcommand{\1}{\boldsymbol{1}}
\newcommand{\bg}{\boldsymbol{g}}
\newcommand{\kc}{\mathcal{K}^n}
\newcommand{\kkc}{\boldsymbol{\mathcal{K}}^n}

\newcommand{\Kp}{K'}
\newcommand{\bKp}{\boldsymbol{K'}}
\newcommand{\Egx}{\E_{\bpi(0)}\left[\e^{\tf \g \cdot \A[z]} \right]}
\newcommand{\Egz}{\E_{\bpi(0)}\left[\e^{\tf \g \cdot \A[z]} \right]}
\newcommand{\gAx}{\tf \g \cdot \A[z]}

\newcommand{\peri}{T}
\newcommand{\tf}{n\peri}
\newcommand{\scgf}{\phi}

\newcommand{\bQ}{\boldsymbol{\overleftarrow{\mathcal{Q}}}}
\newcommand{\uP}{\mathds{P}}
\newcommand{\uPk}{\mathds{P}_{\kk,\bpi(0)}[z]}
\newcommand{\uPkcond}{\mathds{P}_{\kk,\bpi(0)}\left[z \mid \A\left[ z \right] = \textbf{a} \right]}
\newcommand{\uPK}{\mathds{P}_{\K,\bpi(0)}[z]}
\newcommand{\uPkappa}{\mathds{P}_{\bkappa,\bpi(0)}[z]}
\newcommand{\uPmicro}{\mathds{P}^{\text{micro}}_{\textbf{a},\bpi(0)}[z]}
\newcommand{\uPcano}{\mathds{P}^{\text{cano}}_{\g,\bpi(0)}[z]}
\newcommand{\uPkcano}{\mathds{P}_{\bkappa^{\C}, \frac{\C(0) \circ \bpi(0)}{\C(0) \cdot \bpi(0)}}[z]}

\newtheorem{property}{Property}

\begin{document}
\title{Periodically driven jump processes conditioned on large deviations}
\author{Lydia Chabane}
\affiliation{Laboratoire de Physique Th\'eorique (UMR8627), CNRS, Univ. Paris-Sud, Universit\'e Paris-Saclay, 91405 Orsay, France}
\author{Raphaël Chétrite}
\affiliation{Laboratoire J A Dieudonné, UMR CNRS 7351, Université de Nice Sophia
Antipolis, Nice 06108, France}
\author{Gatien Verley}
\affiliation{Laboratoire de Physique Th\'eorique (UMR8627), CNRS, Univ. Paris-Sud, Universit\'e Paris-Saclay, 91405 Orsay, France}
\date{\today}

\begin{abstract}
	We study the fluctuations of systems modeled by Markov jump processes with periodic generators. We focus on observables defined through time-periodic functions of the system's states or transitions. Using large deviation theory, canonical biasing and generalized Doob transform, we characterize the asymptotic fluctuations of such observables after a large number of periods by obtaining the Markov process that produces them. We show that this process, called driven process, is the minimum under constraint of the large deviation function for occupation and jumps.
\end{abstract}

\maketitle

\section{Introduction}

Large deviation theory aims to predict the decay rate of a probability when increasing a parameter, e.g. number of realizations, time, system size. The stochastic variables we consider are physical observables characterizing the system of interest. Statistical properties of such observables follow then in an asymptotic limit. 
Large deviation theory is widely used in statistical physics, both at and out of equilibrium. At equilibrium, it gives a formal point of view on thermodynamic potentials~\cite{Amann1999, Deuschel1991, Dobrushin1992, Donnelly1987, Lanford1973, Ruelle1999}: state functions such as entropy or free energy are large deviation functions (LDFs) or cumulant generating functions (CGFs), while variational principles follow from a saddle point approximation~\cite{Touchette2009}. 
Crucially, large deviation theory also applies to nonequilibrium systems. Examples include chemical reaction networks for which a Lagrangian/Hamiltonian description has been derived in Ref.~\cite{Lazarescu2019vol151}, multi-scale systems with noise-induced transitions~\cite{Grafke2017_vol2017}, and kinetically constrained models such as the East model~\cite{Jack2013}. Algorithms computing numerically the probability of rare events have also been developed in the framework of large deviation theory~\cite{Grafke2019vol29}. More generally, the concept of large deviations appears in many fields such as dynamics of population~\cite{Chauvin1988, Derrida2016}, finance~\cite{Pham2007} and bio-informatics~\cite{Arratia1990}.

Focusing on systems modeled by Markov jump processes, we are interested in the fluctuations of an observable $\A$ extensive in time that satisfies a large deviation principle. $\A$ may be a physical observable such as heat current, work or entropy production. Like in equilibrium statistical mechanics where rare fluctuations inform on all the thermodynamic states of the same system,
a large deviation of $\A$ for a given process corresponds to a typical value of $\A$ for another process. More precisely, we will consider a process conditioned on a given value of $\A$ and connect this conditioned process to an effective conditioning-free process. Such a connection has been done recently in the stationary case in Ref.~\cite{Chetrite2014, Chetrite2015_vol2015}. In this paper, we extend this result for periodically driven processes and for observables involving time-periodic functions. 

This extension is motivated by the fact that many thermodynamic machines, including engines, operate via cycles or under periodic control. Such machines are experimentally studied nowadays at the fluctuating level~\cite{Blickle2011, Martinez2015, Leigh2015, Martinez2017}. Fluctuations in periodically driven systems modeled by Markov processes with time-periodic transition rates have also attracted interest at the theoretical level~\cite{Barato2018_vol2018, Verley2014}. Besides externally driven systems, spontaneous oscillations exist for systems in stationary nonequilibrium, see for instance the Brusselator model of chemical reactions~\cite{Li2016, Andrieux2008_vol128}, or the three-state model studied in Ref.~\cite{Herpich2018PRX}. Achieving a complete theory on the conditioning of those systems requires first to understand the fluctuations of periodically driven processes. 

In this paper, we start in Section~\ref{sec_def} by defining our periodically driven Markov jump process and the conditioning observable $\A$. Then, we introduce the moment generating and the scaled cumulant generating functions (SCGF) for observable $\A$. Next, we write the spectral elements associated with the one-period propagator for the generating function. 
In Section~\ref{sec_honest}, we address the problem of conditioning a Markov process on rare values of $\A$. This conditioned process corresponds to a microcanonical ensemble of trajectories, i.e. trajectory filtrated on the value of $\A$. However, this process has usually no Markov generator. 
Like in equilibrium statistical mechanics, we use in Section \ref{sec_honest} the canonical ensemble of trajectories (by exponentially biasing each trajectory probability) to build a canonical process which is Markovian. In the limit of a large number of periods, we show that the canonical process becomes the so-called driven process that will appear later as an optimal process for which $\A$ converges in probability to the microcanonical value.
Assuming a unique relation between the canonical bias and the conditioning value $\ba$, the driven process defines the conditioning-free process which is asymptotically equivalent to the microcanonical process. 
In Sec.~\ref{sec_var}, we show from a variational point of view that the driven process is the optimizer of the 2.5 LDF of occupations and transition probabilities under the constraint $\A = \ba$.  Hence, the driven process is the most probable process that reproduces the fluctuation $\ba$. This method is in clear analogy with Jaynes' maximum entropy principle in which entropy is replaced by the 2.5 LDF. 
In Section~\ref{sec_ex}, we conclude this paper by illustrating our results on a periodically modulated two level system conditioned on a current that is differently defined on each part of the period. 

\section{Definitions and Notations} \label{sec_def}

\subsection{Periodically driven Markov jump processes}

We consider a continuous-time Markov jump process defined on a finite state space~\cite{Kampen2007}, $z(t)$ giving the state of the system at time $t$. The generator of this Markov process is $\kk$, where $k_{xy} \doteq k_{xy}(t)$ is the transition rate from state $y$ to state $x$ at time $t$. The rates $k_{xy}$ for $x \neq y$ are non-negative and $k_{yy} \doteq -\lambda_y$ where 
\begin{equation} \label{escape_rate}
\lambda_y(t) \doteq \sum_{x \neq y} k_{xy}(t)
\end{equation}
is the escape rate from state $y$ at time $t$. We assume that the transition rates are time-periodic with period $T$: $k(t+\peri) = k(t)$, and denote by $\pi_x \doteq \pi_x(t)$ the probability to be in state $x$ at time $t$. It satisfies the master equation
\begin{equation} \label{Eq_maitresse_standard}
\frac{\partial \pi_x}{\partial t} = \sum_y k_{xy} \pi_y.
\end{equation}
The norm of the probability is conserved by the master equation since by construction $\sum_x k_{xy} = 0, ~ \forall y$: we say that $\kk$ generates an \textit{honest} Markov process. 
The master equation and the normalized initial probability ensure the normalization of the probability at all times. We suppose that at time $t = \tf$ and for $n \rightarrow \infty$, $\bpi$ reaches a periodic solution of the master equation $\bpi^{\text{Tips}}$ --- where TiPS stands for \textit{Time Periodic State} --- i.e. $\bpi^{\text{Tips}}(\tau+\peri) = \bpi^{\text{Tips}}(\tau)$, $\forall \tau \in [0,\peri]$.  

We call a path the succession of states visited by the system in addition to the knowledge of the times at which transitions occur, and in some cases the mechanism by which a new state is reached (for instance, the heat bath the system exchanges energy with when a transition occurs). In the following, we denote a path by $[z]$ and assume that it starts at $t=0$ and ends at $t=\tf$, if not otherwise stated. We label $\{z_i\}_{i=0}^{N}$ the visited states and $\{t_i\}_{i=0}^N$ the times at which the system jumps such that
\begin{eqnarray}
z(t) = z_i \qquad \mathrm{for} \qquad  t_i \leq t < t_{i+1}.
\end{eqnarray}
Time $t_0=0$ is the initial time and $t_N$ is the last jump time before the final time $\tf$. The path probability $\uPk$ of path $[z]$ is given by
\begin{equation} \label{path_proba}
\uPk = \pi_{z_0}(0) \exp\left[ \sum_{i=0}^{N-1} \ln\left(k_{z_{i+1},z_i}(t_{i+1}) \right)  - \int_0^{\tf} \lambda_{z(t)}(t) \ud t \right],
\end{equation}
where the index $\bpi(0)$ refers to the initial state probability.

\subsection{Observable, scaled cumulant generating function and tilted matrix} \label{scgf_results}

We are now interested in the fluctuations of an observable $\A_{\tf} \doteq \A_{\tf}[z]$ that is a real functional of the paths up to the final time $\tf$. For sake of generality, we consider the following two components observable $\A_{t}[z]$ on the shorter time interval $[0,t]$ defined by
\begin{equation} \label{observable}
\A_{t}[z] \doteq \frac{1}{t} \left( \begin{array}{c}
\displaystyle \sum_{i=0}^{N-1}  \, g_{z_{i+1},z_i}(t_{i+1}) \theta(t-t_{i})  \\ \displaystyle \int_0^{t} \ud \tau f_{z(\tau)}(\tau) 
\end{array} \right)
\end{equation}
where $\theta(t)$ is the Heavyside function, $\bm{g}$ and $\bm{f}$ are time periodic functions with period $T$. 
When specifying the components of matrix $\bm{g}$ and vector $\bm{f}$, observable $\A_{\tf}[z]$ may represent physical quantities.  
For instance, it is the number of jumps per unit time for $g_{xy}=1$ and $f_{y}=0$ for all $x$ and $y $. It is the occupation time in state $x$ if $g_{y,z}=0$ and $f_{y} = \delta_{xy} $ with $\delta$ the Kronecker delta. Last, it is work and heat currents exchanged with the reservoirs if $f_x(t)$ is the partial derivative with respect to time $t$ of the system energy in state $x$ and $g_{xy}(t)$ is energy difference between states $x$ and $y$ at time $t$.
 
We assume that $\A_{\tf}$ satisfies the large deviation principle
\begin{equation}
P(\ba) \underset{n \rightarrow \infty}{\sim} \e^{-\tf I(\ba)},
\end{equation}
where $P(\ba)$ is the probability of the rare event $\{\A_{\tf}[z] = \textbf{a} \mid \textbf{a} \in \R^2 \}$ and $I(\ba)$ its associated LDF or rate function. The LDF describes the exponential decay with time of the probability that $\A_{\tf}$ takes a value different from its typical one. In order to study the fluctuations of $\A_{\tf}$ in the long-time limit, we introduce the generating function for   $\A_t$ 
\begin{equation} \label{GF}
G(\g,t) \doteq  \E_{\bpi(0)}\left[e^{t \, \g  \cdot \A_t[z]} \right],
\end{equation}
where $\E_{\bpi(0)}[\cdot]$ is the path average on $[z]$ with initial probability $\bpi(0)$. The central dot $\cdot$ stands for the scalar product and vector $\g = \left( \begin{array}{cc} \gamma_1 & \gamma_2 \end{array} \right)^T$ is the Laplace conjugate variable. 
The generating function at time $t$ imposing the final state $z(t) = x$ writes
\begin{equation} \label{GF_x}
G_x(\g,t) \doteq \E_{\bpi(0)}\left[e^{t \, \g  \cdot \A_t[z]} \delta_{x,z(t)} \right].
\end{equation}
The SCGF for $\A_{\tf}$ is defined by
\begin{equation} \label{SCGF}
\phi(\g) \doteq \underset{n \rightarrow \infty}{\lim} \frac{1}{\tf} \ln G(\g,\tf). 
\end{equation} 
The conditioned generating function $G_x$ satisfies the ordinary differential equation~\cite{Speck2005_vol38, Chernyak2006_vola, Verley2012_vol, Verley2013_vol88, Barato2018_vol2018}
\begin{equation} \label{equa_diff_generating_function}
\partial_t G_x(\g,t) = \sum_{y} \kappa_{xy}(\g,t) G_{y}(\g,t),
\end{equation}
where we have introduced the tilted (or dressed) operator $\bkappa$ of components
\begin{equation} \label{tilted_matrix}
\kappa_{xy}(\g,t) \doteq \left\{
      \begin{aligned}
        k_{xy}(t) \e^{\gamma_1 \, g_{xy}(t)}   & & \mathrm{if} ~ x \neq y, \\
        - \lambda_x(t) + \gamma_2 \, f_x(t) & & \mathrm{if} ~ x = y.
      \end{aligned}
    \right.
\end{equation}
Notice that by definition $\bkappa(\g,\tau+T) = \bkappa(\g,\tau), ~ \forall \tau \in [0,\peri]$. In the following, we keep in mind that $\bkappa$ depends on $\g$ and drop $\g$ in the notations for clarity. The tilted matrix can be seen as the generator of a new process, called the tilted process, that contains information on the large deviations of $\A_{\tf}$, but that is not honest. We formally solve Eq.~\eqref{equa_diff_generating_function} with initial condition $\bm{G}(\g,0) = \bpi(0)$ writing 
\begin{equation} \label{G_x(t)}
\bm{G}(\g,t) =   \bQ_{\bkappa}(t,0) \bpi(0),
\end{equation} 
in term of the propagator
\begin{equation} \label{propagateur}
\bQ_{\bkappa}(t,t_0) \doteq \overleftarrow{\exp} \int_{t_0}^t \bkappa(t') \, \ud t',
\end{equation}
involving the time-ordered exponential $\overleftarrow{\exp}$, see Appendix~\ref{app_propagator}. $\bQ_{\bkappa}(t,t_0)$ is solution of Eq.~\eqref{equa_diff_generating_function}, but
with initial condition $\bQ_{\bkappa}(t_{0},t_0) = \mathbb{1}$ the identity matrix in the state space.

\subsection{Spectral elements of the one-period propagator for the tilted process} \label{spectral_prop}

From now on, we assume that the final time is always $\tf$ and omit the subscript $\tf$ for our generic observable $  \A[z] \doteq \A_{\tf}[z]$. This observable evaluated for a stochastic process $z(t)$ on $[0,\tf]$ becomes the random variable $\A$.
In this section, we relate generating functions for $\A$ to the spectrum and eigenspace of the propagator $\bQ_{\bkappa}(T,0)$,  giving them a physical interpretation.

Let $\rho_T$ be the highest eigenvalue of $\bQ_{\bkappa}(\peri,0)$, $\ff_T$ its right (column) eigenvector and $\h_T$ its left (row) eigenvectors:
\begin{eqnarray}
\bQ_{\bkappa}(\peri,0) \ff_T &=& \rho_T \, \ff_T, \label{right_eig}\\
\h_T \bQ_{\bkappa}(\peri,0) &=& \rho_T \, \h_T. \label{left_eig}
\end{eqnarray}
The eigenvectors $\h_T$ and $\ff_T$ can be chosen up to a multiplicative constant that we set by imposing 
\begin{eqnarray}
 \1 \cdot \ff_T & = & 1, \label{phi_normal} \\
\h_T \cdot \ff_T  & = & 1, \label{pdriven_normal}
\end{eqnarray}
where $\1$ is the vector whose components are all $1$. Furthermore, we remark that $\h_T \cdot \bpi(0) < \infty$ since the state space is finite.

We now make a connection between the spectral elements of the propagator and the large deviations of $\A$. From Eqs.~(\ref{GF_x}, \ref{G_x(t)}) and using the periodicity of $\bkappa$, the generating functions $G$ and $G_x$ at time $\tf$ write
\begin{eqnarray}
G(n\peri) &=& \E_{\bpi(0)} \left[e^{\gAx} \right] = \1 \cdot \left( \bQ_{\bkappa}(\peri,0)^n \bpi(0) \right) = \sum_{x,y} \left[ \overleftarrow{\mathcal{Q}}_{\bkappa}(\peri,0)^n\right]_{xy} \pi_y(0), \label{defG} \\
G_x(n\peri) &=& \E_{\bpi(0)} \left[e^{\gAx} \delta_{x,z(\tf)} \right] = \sum_{y} \left[ \overleftarrow{\mathcal{Q}}_{\bkappa}(\peri,0)^n\right]_{xy} \pi_y(0). \label{defGx}
\end{eqnarray} 
The asymptotic expansion of $\bQ_{\bkappa}(\tf,0)$ at large~$n$ is given by
\begin{equation} \label{dev_assyp_Q}
\bQ_{\bkappa}(\peri,0)^n \underset{n \rightarrow \infty}{\approx} (\rho_T)^n \, \ff_T \, \h_T.
\end{equation}
With Eqs.~(\ref{phi_normal}--\ref{defG}), it yields
\begin{equation}
\scgf(\g) = \lim_{n \rightarrow \infty} \frac{1}{\tf} \ln \Egz = \frac{1}{\peri} \ln \rho_T. \label{scgf_plus_grande_vp}
\end{equation}
The SCGF $\scgf$ is proportional to the logarithm of the highest eigenvalue of the single-period propagator $\bQ_{\bkappa}(\peri,0)$~\cite{Verley2013_vol88, Barato2018_vol2018}. Similarly, combining Eqs.~(\ref{defG},\ref{defGx},\ref{dev_assyp_Q}) and using Eqs.~(\ref{phi_normal}--\ref{pdriven_normal}), we find
\begin{eqnarray}
\lim_{n \rightarrow \infty} \e^{-\tf \scgf} \E_{x_0}\left[\e^{\gAx} \right] & = & (l_{T})_{x_0}, \label{hpath} \\
\lim_{n \rightarrow \infty} \frac{\E_{\bpi(0)}\left[\e^{\gAx} \delta_{z(\tf),x} \right]}{\Egz} & = & (r_T)_{x}, \label{phipath}
\end{eqnarray}
where $\E_{x_0}$ is the path average over $[z]$ with a Dirac delta centered on $x_0$ as initial probability. Hence, Eqs. (\ref{scgf_plus_grande_vp}, \ref{hpath}, \ref{phipath}) allow to write the eigenvectors of the propagator in term of path averages. Hence, we have extended for periodically driven processes the results of Ref.~\cite{Chetrite2014} that hold in the stationary case.

\section{Conditioning: building an effective process} \label{sec_honest}

We saw in the previous section that the propagator based on the tilted matrix allows to describe the large deviations of $\A$. We are now interested in conditioning our original Markov process by filtering the ensemble of paths to select those leading to a chosen value of $A$. This defines the so-called microcanonical process for which we aim to find an equivalent Markov process in the long-time limit. Some results of this section are stated in Section~$8.8$ of Ref.~\cite{ChetriteHDR}.

\subsection{Microcanonical process}

The process $z(t)$ conditioned on the event $\{\A[z] = \textbf{a} \mid \textbf{a} \in \R^2 \}$ is described by the microcanonical path probability~\cite{Chetrite2014}
\begin{equation} \label{pathprobamicro}
\uPmicro = \uPkcond.
\end{equation}
In general, there is no Markov generator that can generate exactly this microcanonical ensemble of path. Yet, there is another process called the canonical process that is Markovian and that has the interesting property to be asymptotically equivalent (in a way to be defined later) to the microcanonical process~\cite{Chetrite2014, Touchette2015}.

\subsection{Canonical process} \label{sec_microcano}

The canonical path probability is connected to the original process by an exponential tilting of the path probability $\uPk$~\cite{Chetrite2014}:
\begin{equation} \label{pathprobacano}
\uPcano \doteq \frac{\e^{\gAx} \uPk}{\Egx} = \frac{\uPkappa}{\Egx},
\end{equation}
where 
\begin{equation} \label{path_proba_kappa}
\uPkappa = \uPk \e^{\gAx}
\end{equation}
is obtained by injecting Eq.~\eqref{tilted_matrix} in the definition of a path probability of Eq.~\eqref{path_proba}. This path probability is a natural generalization at the path level of the equilibrium probability in the canonical ensemble. This definition has already been used in many articles, for instance for the simulation of transition paths associated with glassy systems~\cite{Garrahan2009, Binder2006}.

The canonical process is honest and Markovian as it is generated by a Markov generator that we explicit in the following. To do so, we look for a generator $\kkc$ that satisfies
\begin{equation}
\uP_{\kkc, \bpi'(0)}[z] \doteq \uPcano,
\end{equation}
where $\bpi'(0)$ is an initial probability that may be different from $\bpi(0)$. Moreover, we want $\kkc$ to generate an honest process, hence we look for a vector $\C$ such that
\begin{equation} \label{doob_kcano}
\kkc \doteq \bkappa^{\C} \doteq \mathcal{D}(\C) \bkappa \, \mathcal{D}(\C)^{-1} - \mathcal{D}(\C)^{-1} \mathcal{D}(\C \bkappa)
\end{equation}
is built from a Doob transform as defined in Appendix~\ref{Doob}. Looking at the path probability in Eq.~(\ref{path_proba_doob}) obtained from a Doob transform, $\C \doteq \C(t)$ should be chosen as the solution of
\begin{eqnarray}
\begin{cases}
\dot{\C} = - \C \bkappa, \\
\C(\tf) = \1,
\end{cases}
\end{eqnarray}
such that the time extensive term in the exponential of Eq.~(\ref{path_proba_doob}) vanishes. Equivalently, using property~\ref{P3} of Appendix~\ref{Properties}, $\C$ writes
\begin{equation} \label{Cn}
\C(t) = \1 ~ \bQ_{\bkappa}(\tf,t).
\end{equation}

Injecting Eq.~(\ref{doob_kcano}) in Eq.~(\ref{path_proba_doob}), we obtain that the path probability associated with $\bkappa^{\C}$ is given by
\begin{equation}
d \mathds{P}_{\bkappa^{\C},\bpi(0)}[z] = d \mathds{P}_{\bkappa,\bpi(0)}[z] \, (C^n_{x_0})^{-1}(0),
\end{equation}
or equivalently
\begin{equation}
d \mathds{P}_{\bkappa,\bpi(0)}[z] = d \mathds{P}_{\bkappa^{\C}, ~ \bpi(0) \circ \C(0)}[z],
\end{equation}
where $\circ$ is the Hadamard product: $( \bm{u} \circ \bm{v} )_x \doteq u_x v_x$. From Eqs.~(\ref{defG}) and (\ref{Cn}), we remark that the generating function can be expressed in term of $\C$ as
\begin{equation}
G(\tf) = \C(0) \cdot \bpi(0),
\end{equation}
implying that the normalizing factor of the canonical path probability can be used to normalize the initial condition vector in the path probability for the Doob transform of $\bkappa$ 
\begin{equation}
\uPcano = \uPkcano.
\end{equation}
In other words, the canonical path probability is associated with the generator $\kkc = \bkappa^{\C}$ for the initial probability~$\C(0) \circ \bpi(0) / [\C(0) \cdot \bpi(0)]$. 
This shows that the canonical process  has a corresponding Markov generator and can thus be considered as a Markov process.

In the next section, we focus on the asymptotic dynamics in the limit $n \rightarrow \infty$ by considering the process towards which the canonical process converges at long time.

\subsection{Driven process} \label{direct_spectral_approach}

The driven process is defined as the limit of the canonical process as $n \rightarrow \infty$~\cite{Chetrite2014}. Since the canonical process comes from the Doob transform of the tilted operator using $\C$, the driven process will be built similarly.
In the limit $n \rightarrow \infty$, using Eqs.~(\ref{phi_normal}, \ref{dev_assyp_Q}), we find that $\C(\tau)$ for $\tau \in [0,T[$ is given asymptotically by
\begin{eqnarray} \label{dev_Cn}
\C(\tau) = \1 ~ \bQ_{\bkappa}(\peri,0)^{n} \left[ \bQ_{\bkappa}(\tau,0) \right]^{-1} \underset{n \rightarrow \infty}{\sim} (\rho_T)^n \h_T \left[ \bQ_{\bkappa}(\tau,0) \right]^{-1}.
\end{eqnarray}
Since scalar constants play no role in the Doob transform, we introduce the function of time $\h \doteq \h(\tau)$:
\begin{equation} \label{h_def}
\h(\tau) \doteq \h_T \left[ \bQ_{\bkappa}(\tau,0) \right]^{-1},
\end{equation} 
that is by construction the solution of
\begin{equation} \label{h_equadiff}
\begin{cases}
\dot{\h} = - \h \bkappa, \\
\h(0) = \h_T.
\end{cases}
\end{equation}
Using Eq.~\eqref{left_eig} and the periodicity of $\bkappa$, the vector $\h$ satifies 
\begin{equation} \label{prop_h(t+T)}
\h(\tau+\peri) = \rho_T^{-1} \h(\tau).  
\end{equation}
We define the Markov generator $\K \doteq \K(\g,\tau)$ of the driven process at all time $\tau$ by the Doob transform of the tilted matrix $\bkappa$ associated with vector $\h$:
\begin{equation} \label{Doob transform_Kh}
\K \doteq \bkappa^{\h} = \mathcal{D}(\h) \bkappa \, \mathcal{D}(\h)^{-1} -\mathcal{D}(\h)^{-1} \mathcal{D}(\h \bkappa).
\end{equation}
Note that the positivity of $\h(t)$ at all $t$ is ensured by the positivity of $\bQ_{\bkappa}(t,0)$ and Perrond-Frobenius theorem. From Eqs. \eqref{doob_kcano} and \eqref{dev_Cn}, we see that the generator of the driven process is given by the limit of the canonical transition matrix as $n \rightarrow \infty$:
\begin{equation} \label{kcano_K_lim}
\lim_{n \rightarrow \infty} \kkc(\tau) = \bkappa^{\h}(\tau) = \K(\tau).
\end{equation}
One interesting property of $\K$ is its periodicity. Indeed, from Eq.~\eqref{prop_h(t+T)} and the periodicity of $\bkappa$, we have $\K(\tau+\peri) = \K(\tau), ~ \forall \tau \in [0,T]$.

In the following, we show that the driven and canonical path probabilities are asymptotically equivalent. Two paths $\mathds{P}_n$ and $\mathds{Q}_n$ are said to be logarithmically equivalent if $\lim_{n \rightarrow \infty} \frac{1}{n} \ln  \frac{\mathds{P}_n}{\mathds{Q}_n}$, and we denote it $\mathds{P}_n \asymp \mathds{Q}_n$. In this case, if an observable satisfies a large deviation principle with respect to $\mathds{P}_n$ and $\mathds{Q}_n$, then the corresponding LDFs vanish at the same values. This means that, in case of logarithmic equivalence, this observable takes the same typical values with respect to both paths in the limit $n \rightarrow \infty$~\cite{Chetrite2014, Touchette2015}. Using Eqs.~(\ref{scgf_plus_grande_vp}, \ref{h_equadiff}, \ref{Doob transform_Kh}) and \eqref{path_proba_doob}, the path probability of the driven process writes
\begin{equation} \label{pathproba_K}
\uPK = \uPkappa \, l_{x_{\tf}}(0) \, \e^{-\tf\scgf} \, l^{-1}_{x_0}(0).
\end{equation}
Using the definitions of the canonical path probability \eqref{pathprobacano} and driven path probability \eqref{pathproba_K}, we get
\begin{equation}
\frac{\uPK}{\uPcano} = l_{x_{\tf}}(0) \, \e^{-\tf\scgf} \, l_{x_0}(0) \, \Egx.
\end{equation} 
Hence, using the definition of the SCGF \eqref{SCGF}, we finally find:
\begin{equation}
\lim_{n \rightarrow \infty} \frac{1}{\tf} \ln \frac{\uPK}{\uPcano} = 0.
\end{equation}
The driven path probability and the canonical path probability are then logarithmically equivalent:
\begin{equation} \label{eq_driven_cano}
\uPK \asymp \uPcano.
\end{equation}

Finally, we remark that the TiPS probability for the driven process can be obtained from the solution of the initial value problem of Eq.~(\ref{h_equadiff}) giving for $\h$ and the intial value problem for~$\ff\doteq \ff(\tau)$
\begin{equation} \label{phi_equadiff}
\begin{cases}
\dot{\ff} = \bkappa \ff, \\
\ff(0) = \ff_T,
\end{cases}
\end{equation}
or alternatively
\begin{equation}
\ff(t) \doteq \bQ_{\bkappa}(t,0) \ff_T.
\end{equation} 
Using Eq.~\eqref{right_eig} and the periodicity of $\bkappa$, the vector $\ff$ satisfies:
\begin{equation} \label{prop_f(t+T)}
\ff(t+\peri) = \rho_T \ff(t).  
\end{equation}
The TiPS probability of the driven process $\bmu \doteq \bmu(t)$, defined as the $\peri$-periodic solution of the master equation:
\begin{equation} \label{driven_proba}
\begin{cases}
\frac{\ud \bmu}{\ud t} = \K \bmu \\
\bmu(0) = \bmu(\peri),
\end{cases}
\end{equation}
writes in term of the vectors $\h$ and $\ff$
\begin{equation} \label{tips_driven}
\bmu(t) = \h(t) \circ \ff(t).
\end{equation}
Indeed, Eqs.~(\ref{h_equadiff}, \ref{Doob transform_Kh}, \ref{phi_equadiff}) yield for all $x$
\begin{eqnarray}
\sum_y K_{xy}(l_y r_y) & = & \sum_y \Big\{l_x \kappa_{xy} l_y^{-1} l_y r_y - l_x^{-1} l_{y} \kappa_{yx} l_x r_x  \Big\} \\
& = & l_x \dot{r}_x + \dot{l}_x r_x \\
& = & \frac{\ud}{\ud t} (l_x r_x),
\end{eqnarray} 
while Eqs.~(\ref{prop_h(t+T)},\ref{prop_f(t+T)}) lead to
\begin{eqnarray}
\h(0) \circ \ff(0) & = & \h(T) \circ \ff(T),
\end{eqnarray}
which proves that $\bmu$ is the solution of Eq.~(\ref{driven_proba}). Notice that our normalization choice in Eq.~(\ref{pdriven_normal}) ensures the normalization of $\bmu (0)$. We emphasize that in this section we have essentially extended for the periodic case the results of~\cite{Chetrite2014} for the stationary case.

\subsection{Free-conditioning process}

We saw that the canonical process tends to the driven process in the long-time limit. We still need to obtain the Markov process  that is equivalent at large time to the microcanonical process for which $\{\A[z]=\textbf{a}~\mid~\textbf{a}~\in~\R^2~\}$. 
This process is called the \textit{free-conditioning process}. 

From Ref.~\cite{Touchette2015}, the canonical and microcanonical path probabilities (thought respectively as biased and conditioned path ensembles based on $\uPk$ and observable $\A$) are logarithmically equivalent if the LDF $I$ is convex at $\ba$.
In this case, and assuming that $I$ is differentiable for simplicity, the equivalence holds for $\g = \nabla I(\ba)$, where $\nabla I(\ba)$ is the gradient of $I$ evaluated at $\ba$. Mathematically, this writes:
\begin{equation} \label{eq_micro_cano}
\uPmicro \asymp \uPcano \left|_{\g = \nabla I(\ba)} \right.
\end{equation}
When combined with the logarithmic equivalence between the driven and canonical path probabilities of Eq.~(\ref{eq_driven_cano}), we find that the free-conditioning process is the driven process for $\g = \nabla I(\ba)$. Mathematically, this writes:
\begin{equation}
\uPmicro \asymp \uPK \left|_{\g = \nabla I(\ba)} \right.
\end{equation}
Notice that if $I$ is not convex at $\ba$, there is no Markov process equivalent to the microcanonical process.

\section{Variational representation of the driven process} \label{sec_var}

In this section, we derive the driven process from a variational approach. This route requires to determine the functional to be minimized and playing the role of entropy in Jaynes' maximum entropy principle of statistical mechanics. This functional is the LDF for occupations and transition probabilities. We find that the driven process is the most probable process for which observable $\A$ takes asymptotically a chosen value.

The \textit{occupation density} $p^n_{x}(\tau)[z]$ at phase $\tau \in [0, \peri[$ is the path functional 
\begin{equation} 
p^n_{x}(\tau)[z] = \frac{1}{n} \sum_{m=0}^{n-1} \delta_{x, z(\tau + m\peri)}
\end{equation}
that counts the fraction of time the system has been in state $x$ at phase $\tau $ of each period along the path $[z]$. The  occupation density is a positive vector of norm $1$ that converges to $\bpi^{\text{Tips}}$, when considering the process of generator $\kk$. The \textit{empirical transition probability} $\omega^n_{xy}(\tau)[z]$
\begin{equation}
\omega^n_{xy}(\tau)[z] = \frac{1}{n} \sum_{m=0}^{n-1} \frac{1}{\ud \tau} \sum_{s \in [\tau,\tau+\ud \tau]} \delta_{y, z(s^- + m\peri)} \delta_{x, z(s^+ + m\peri)},
\end{equation}
with $\ud \tau$ an infinitesimal time, measures the number of transitions $y \rightarrow x$ per unit of time at phase $\tau$, or more precisely during $[\tau,\tau+d\tau]$. When considering the process of generator $\kk$, $\om$ converges to $\kk \circ \bpi^{\text{Tips}}$, with $\left[\kk \circ \bpi^{\text{Tips}} \right]_{xy} \doteq k_{xy} \pi^{\text{Tips}}_y$.  We can rewrite the conditioning observable $\A$ of Eq.~\eqref{observable} in terms of $\p^n[z]$ and $\om^n[z]$ using the periodicity of $\bm{f}$ and $\bm{g}$: 
\begin{equation} \label{observabletest}
\A\left(\om^n[z], \p^n[z]\right) = \left( \begin{array}{c} A_1(\om^n[z]) \\ A_2(\p^n[z]) \end{array} \right) \doteq \left( \begin{array}{c}
\frac{1}{\peri} \int_0^{\peri} \ud \tau  \sum_{x, y \neq x} \, \omega^n_{xy}[z](\tau) ~ g_{xy}(\tau) \\ \frac{1}{\peri} \int_0^{\peri} \ud \tau \sum_x p^n_x[z](\tau) ~ f_x(\tau)
\end{array} \right).
\end{equation}

At long time $\tf$, the probability to observe the  occupation $\p^n[z] = \p$ and the empirical transition probability $\om^n[z] = \om$ satifies a large deviation principle:
\begin{equation}
P(\om,\p) \underset{n \rightarrow \infty}{\sim} \e^{\tf I_{2.5}(\om, \p)},
\end{equation}
where the 2.5 LDF is given by~\cite{Bertini2018_vol19}
\begin{equation} \label{LDF_omega_p}
I_{2.5}(\om, \p) =  \sum_{y, x \neq y}  \frac{1}{\peri} \int_0^\peri \ud \tau \left[ \; p_y(\tau) ~ \left( k_{xy}(\tau) - \frac{\omega_{xy}(\tau)}{p_y(\tau)} \right) + \omega_{xy}(\tau) \ln \frac{\omega_{xy}(\tau)}{k_{xy}(\tau) p_y(\tau)} \right],
\end{equation} 
with $\p(0) = \p(T)$ and $\om(0) = \om(\peri)$. This expression holds only for conservative transition probabilities with $\dot{p}_x(\tau)=\sum_y(\omega_{xy}(\tau)-\omega_{yx}(\tau))$, $\forall x$, and normalized occupations with $\sum_y p_y(\tau) = 1$, otherwise $I_{2.5}$ is infinite. Notice that $I_{2.5}(\om, \p)$ vanishes for $\p = \bpi^{\text{Tips}}$ and $\om = \kk \circ \bpi^{\text{Tips}}$. Hence, without conditioning, $\A$ converges to $\A(\kk \circ \bpi^{\text{Tips}}, \bpi^{\text{Tips}})$ as $n \rightarrow \infty$.

As before, we are interested in conditioning our process on the event $\{ \A(\om,\p) = \ba \mid \ba \in \R^2 \}$. We look for the most probable pair $(\om, \p)$ compatible with $\A(\om,\p) = \ba$. This pair coincides with the mean value of $(\om^n[z], \p^n[z])$ under the microcanonical path probability~\cite{Chetrite2015_vol2015, Touchette2009}. It is obtained by minimizing the 2.5 LDF under the following constraints
\begin{itemize}
\item \textbf{C0:} $\qquad \A(\om,\p) = \ba,$
\item \textbf{C1:} $\qquad \sum_y p_y(\tau) = 1, \forall \tau \in [0,\peri]$,
\item \textbf{C2:} $\qquad \dot{p}_x(\tau) = \sum_y (\omega_{xy}(\tau) - \omega_{yx}(\tau)),  \forall x \textrm{ and } \forall \tau \in [0,\peri]$,
\item \textbf{C3:} $\qquad \p(T) = \p(0)$,
\item \textbf{C4:} $\qquad \om(T) = \om(0)$.
\end{itemize}
This optimization problem amounts to compute the LDF of $\A$ and writes mathematically
\begin{equation} \label{I(a)}
I(\ba) = \underset{\p,\om|\{\textbf{Ci}\}_{i=0}^4}{\text{inf}} \Big\{I_{2.5}(\om,\p) \Big\}.
\end{equation}
Eq.~\eqref{I(a)} is known as the \textit{contraction principle}. Intuitively, the optimizer $(\om_{\ba},\p_{\ba})$ is expected to be associated with the generator of the conditioning-free process or similarly the microcanonical process as $n \rightarrow \infty$. Equivalently, we can instead compute the SCGF given by the Legendre tranform of the LDF in Eq.~\eqref{I(a)}:
\begin{equation} \label{scgf_legendre_lagrange}
\phi(\g) = \underset{\p,\om|\{\textbf{Ci}\}_{i=1}^4}{\text{sup}} \left\{\g \cdot \A(\om,\p) - I_{2.5}(\om,\p) \right\}.
\end{equation}
The solution $(\om_{\g},\p_{\g})$ of Eq.~\eqref{scgf_legendre_lagrange} is the typical value of $(\om^n[z], \p^n[z])$ under the canonical path probability~\cite{Chetrite2015_vol2015, Touchette2009}. It is expected to be associated with the generator of the driven process. For convex LDF, Eqs.~$\eqref{I(a)}$ and $\eqref{scgf_legendre_lagrange}$ have the same solutions, i.e. if $I$ is convex at $\ba$, $(\om_{\ba},\p_{\ba}) = (\om_{\g},\p_{\g})$ for $\g = \bm{\nabla} I(a)$. This is in agreement with the equivalence of the microcanonical process and the conditioned-free process/driven process for $\g = \bm{\nabla} I(a)$~\cite{Chetrite2015_vol2015}. In the following, we recover this result through direct calculation of the optimum of

\begin{eqnarray} \label{Lagrange_func}
\mathcal{F}(\om,\p) & = & - I_{2.5}(\om,\p) + \gamma_1 ~ A_1(\om) + \gamma_2 ~ A_2(\p)  \nonumber \\
& - & \frac{1}{\peri} \int_0^\peri \ud \tau ~ c(\tau) ~  \left[~\sum_y p_y(\tau) - 1 ~\right] \nonumber \\
& - & \frac{1}{\peri} \int_0^\peri \ud \tau \sum_x u_x(\tau) ~ \left[~ \dot{p}_x(\tau) - \sum_y \left(\omega_{xy}(\tau) - \omega_{yx}(\tau)\right)~\right],
\end{eqnarray}
where $c$ and $\uu$ are time dependent Lagrange multipliers respectively associated with the constraints \textbf{C1} and \textbf{C2}. We assume in addition that $\uu(T) = \uu(0)$. Functional derivatives with respect to occupation and transition probabilities yield
\begin{eqnarray} \label{contraction}
\left\lbrace
\begin{array}{ccc}
\frac{\partial \mathcal{F}}{\partial \omega_{xy}(\tau)} = 0 & \Rightarrow & \ln \frac{\omega_{xy}(\tau)}{k_{xy}(\tau) p_y(\tau)} + (u_y(\tau) - u_x(\tau)) - \gamma_1 \, g_{xy}(\tau) = 0 \qquad \mathrm{for} ~ x \neq y,   \\
\frac{\partial \mathcal{F}}{\partial p_{y}(\tau)} = 0 & \Rightarrow & \sum_{x \neq y} \bigg[~k_{xy}(\tau) - \frac{\omega_{xy}(\tau)}{p_{y}(\tau)} ~ \bigg] + \, c(\tau) - \gamma_2 \, f_y(\tau) ~ - \dot{u}_y(\tau) = 0. 
\end{array} \right.
\end{eqnarray}
We transform the first equation of \eqref{contraction} into
\begin{equation} \label{contraction_om2}
\omega_{xy}(\tau) = \Kp_{xy}(\tau) ~ p_y(\tau),
\end{equation}
with
\begin{equation} \label{K} 
\Kp_{xy}(\tau)  \doteq  k_{xy}(\tau) ~ \e^{\gamma_1 g_{xy}(\tau)} ~ \e^{u_{x}(\tau) - u_{y}(\tau)} = \kappa_{xy}(\tau) \e^{u_{x}(\tau) - u_{y}(\tau)},
\end{equation} 
for $x \neq y$ and $\tau \in [0,\peri]$. We define the diagonal elements such that the sum over the lines of any column of $\bKp(\tau)$ vanishes so that $\bKp$ satisfies $\dot{\p} = \bKp \p$ via condition \textbf{C2}:
\begin{equation}
\Kp_{yy}(\tau)  \doteq  - \sum_{x \neq y} \Kp_{xy}(\tau) \doteq ~ - ~ \Lambda'_{y}(\tau).
\end{equation}
From condition \textbf{C3}, $\p$ is the TiPS probability associated with $\bKp$. 
As suggested by the notation, $\bKp$ will turn out to be the generator $\K$ of the driven process defined in \eqref{Doob transform_Kh}. The second equation of \eqref{contraction} becomes
\begin{equation} \label{opt_p}
c(\tau) = \sum_{x \neq y} K'_{xy} - \lambda_{y}(\tau) + \gamma_2 \, f_y(\tau) + \dot{u}_y(\tau), \qquad  \forall y.
\end{equation}
Using (\ref{tilted_matrix}, \ref{K}), we get
\begin{equation} \label{opt_p2}
c = \sum_{x \neq y} \kappa_{xy} \e^{u_x-u_y} + \kappa_{xx} + \dot{u}_y.
\end{equation}
Multiplying by $e^{u_y}$, we finally obtain
\begin{equation} \label{equa_diff_eu}
\begin{cases}
\frac{\ud}{\ud t}\left(e^{\uu}\right) = - (e^{\uu}) \left(\bkappa - c \mathds{1} \right), \\
e^{\uu(0)} = e^{\uu(\peri)},
\end{cases}
\end{equation}
with $(\e^{\uu})_x \doteq \e^{u_x}$. The formal solution of \eqref{equa_diff_eu} writes:
\begin{eqnarray}
\e^{\uu(t)} = \e^{\uu(T)} \bQ_{\bkappa - c \mathds{1}}(\peri,t) = \e^{\uu(0)} \e^{-\int_t^\peri c}~ \bQ_{\bkappa}(\peri,t),
\end{eqnarray}
where we used Property \ref{P5} of Appendix~\ref{Properties} in the second equality. Taking $t=0$:
\begin{equation} \label{expu_vp}
\e^{\uu(0)} \bQ_{\bkappa}(\peri, 0) = \e^{\int_0^\peri c}~\e^{\uu(0)}.
\end{equation} 
Hence the optimization with respect to $\p$ leads to a spectral equation. Since the vector $\e^{\uu(0)} $ has positive components, by Perron-Frobenius theorem it is the unique --- up to a multiplicative constante --- left eigenvector of $\bQ_{\bkappa}(\peri, 0)$ associated with its largest eigenvalue $\e^{\int_0^\peri c} = \rho_T$. From Eq.~\eqref{scgf_plus_grande_vp}, we find that the SCGF writes $\phi = \frac{1}{\peri} \int_0^\peri c$. We recover this result in Appendix \ref{app_scgf} directly from Eq.~\eqref{scgf_legendre_lagrange}. Notice that we can rewrite Eq.~\eqref{equa_diff_eu} as:
\begin{equation} \label{eq_diff_h_var}
\frac{\ud}{\ud t}\left(e^{\uu + \int_t^T c}\right) = - (e^{\uu +\int_t^T c}) \bkappa \\
\end{equation}
Hence the vector $e^{\uu + \int_t^T c}$ is solution of
\begin{equation}
\begin{cases}
\dot{\X} = - \X \bkappa, \\
\X(T) = \rho_T^{-1} \X(0).
\end{cases}
\end{equation}
From Eq.~\eqref{h_equadiff}, we conclude that the vector $\h$ that appears in the Doob transform leading to the driven generator is related to the Lagrange multipliers through
\begin{equation} \label{h_def_var}
\h(t) = e^{\uu(t) + \int_t^T c}.
\end{equation} 
We emphasize that $\uu(t)$ is set up to an additive and time-dependent function constant in the state space. Indeed, if \textbf{C2} is satisfied for all states but one then it is satisfied for all states (in view of \textbf{C1}). Then, Eq.~(\ref{h_def_var}) is a choice for this remaining degree of freedom in $\uu(t)$. 

We now show that the transition rate matrix $\bKp$ generates the driven process, i.e. we show that $\bKp$ writes as the Doob transform of $\bkappa$ associated with the vector $\h$. Using Eqs.~(\ref{tilted_matrix}, \ref{opt_p}, \ref{h_def_var}), we transform Eq.~$\eqref{K}$ into
\begin{eqnarray}
\Kp_{xy} & = & \kappa_{xy} \e^{u_{x} - u_{y}} - \left[ \kappa_{xx} + \Lambda'_{x} \right] \, \delta_{xy}, \\
& = & \kappa_{xy} \e^{u_{x} - u_{y}} - \left[ \, \left( - \lambda_x + \gamma_2 f_x \right) + \left( c + \lambda_{x} - \gamma_2 \, f_x - \dot{u}_x \right) \, \right] \, \delta_{xy}, \\
& = & \e^{u_{x}} \kappa_{xy}  \e^{- u_{y}} - ~ \left[ c - \dot{u}_x \right] \, \delta_{xy},\\
& = & l_x \kappa_{xy} l_x^{-1} + l_x^{-1} \dot{l}_x \, \delta_{xy}, \\
& = & l_x \kappa_{xy} l_x^{-1} - l_x^{-1} (\h \bkappa)_x \, \delta_{xy}, \\
&=& \kappa^{\h}_{xy}.
\end{eqnarray}
Hence, we conclude that $\bKp = \K$.
Then, the optimum of Eq.~\eqref{scgf_legendre_lagrange} is reached for $\p = \bmu$ the TiPS probability of the driven process with generator $\K$, and
$\om = \K \circ \p$ the directional probability current associated with the probability $\p$ and rate matrix $\K$:
in the stationary case, we recover the results of~\cite{Chetrite2015_vol2015, Verley2016_vol93}. 

To conclude this section, the driven process is the most probable process that reproduces the dynamics satisfying the imposed value of the conditioning observable. In other words, it is the generator of the Markov process for which the conditioning observable takes asymptotically the imposed value as a mean value.

\section{Illustration on a solvable modulated two-level system} \label{sec_ex}

In this section, we consider a two level system with states denoted by $| \pm \rangle$. For simplicity, the transition rate matrix is chosen symmetric and piecewise-constant. We take as conditioning observable a current defined through a time-periodic function. We compute the SCGF and the rate matrices $ \kkc$ and $\K$ for the canonical and driven processes. We study the convergence of the canonical transition rates toward the driven one as the number of periods $n$ grows. We also comment qualitatively the influence of the conditioning on the transition rates of the driven process.

The transition rate matrix used to model the system writes
\begin{equation}
\kk(t) = 
\left(\begin{array}{cc}
-k(t) & k(t) \\
k(t) & -k(t)
\end{array}\right).
\end{equation}
The rate $k(t)$ is a $T$-periodic and piecewise constant function of time 
\begin{equation}
k(t) = 
\begin{cases}
k^0  \qquad \mathrm{for} ~ t \in [0,\alpha T[, \\
k^1  \qquad \mathrm{for} ~ t \in [\alpha T,T[,
\end{cases}
\end{equation}
where $k^{i}>0$, $i=0,1$ are two constants. We chose $k^{0} = 1 $ to set the time scale.
We use $\alpha \in ]0,1[$ as duty cycle of the piecewise modulation.
The observable $A$ is the scalar path functional 
\begin{equation}
A[z] = \frac{1}{\tf} \sum_{t \in [0,\tf] \, | \, z({t^+}) \neq z({t^-})} \, g_{z(t^+),z(t^-)}(t),
\end{equation}
where we assume that $\bg$ is antisymmetric and piecewise constant with duty cycle $\alpha$, i.e. $g_{-+}(t) = - g_{+-}(t) \doteq g(t)$ and 
\begin{equation}
g(t) = 
\begin{cases}
g^0  \qquad \mathrm{for} ~ t \in [0,\alpha T[, \\
g^1  \qquad \mathrm{for} ~ t \in [\alpha T,T[.
\end{cases}
\end{equation}
When $g^0=1$ and $g^1 =0$ for instance, $A$ counts the net number of transitions $|+ \rangle \rightarrow |-\rangle $ occuring in the first part of each period. With $ |+\rangle $, $|- \rangle$ respectively the first and second basis vectors, the tilted operator associated to $\kk$ and observable $A$ writes
\begin{equation}
\bkappa(\g,t) = 
\left(\begin{array}{cc}
-k(t) & k(t)\e^{-\g g(t)} \\
k(t)\e^{\g g(t)} & -k(t)
\end{array}\right).
\end{equation}

Our theory relies on the propagator $\bQ_{\bkappa}(t,0)$ that we shall now obtain to proceed. The tilted operator being piecewise constant, this propagator writes
\begin{equation}
\bQ_{\bkappa}(t,0) = \e^{t\bkappa} + \Big[ \e^{(t-\alpha T)\bkappa} \e^{\alpha T \bkappa} - \e^{t \bkappa} \Big] \theta(t - \alpha T).
\end{equation}
For $t \in [0,\alpha T$[, we obtain the more explicit expression
\begin{equation} \label{Q0}
\bQ_{\bkappa}(t,0) = \e^{-k^0 t} \left(
\begin{array}{cc} 
\cosh(k^0 t) & \e^{-\g g^0} \sinh(k^0 t) \\
\e^{\g g^0} \sinh(k^0 t) & \cosh(k^0 t)
\end{array} 
\right),
\end{equation}
while for $t \in [\alpha T, T[$, and introducing $t^0 \doteq \alpha T$ and $t^1 \doteq t^{1}(t) \doteq t - \alpha T$, we have
\begin{eqnarray} \label{Q1}
\begin{array}{l}
\bQ_{\bkappa}(t,0) = \bQ_{\bkappa}(t^0 + t^1,0) = \e^{-k^0 t^0-k^1 t^1}  \\ \\ \times  \left(
\begin{array}{cc} 
\prod_{i} \cosh(k^i t^i) + \prod_{i} \e^{\g (1-2i) g^i} \sinh(k^i t^i)  & \sum_{i} \e^{-\g g^i} \sinh(k^i t^i) \cosh(k^{1-i} t^{1-i}) \\
\sum_{i} \e^{\g g^i} \sinh(k^i t^i) \cosh(k^{1-i} t^{1-i}) & \prod_{i} \cosh(k^i t^i) + \prod_{i} \e^{-\g (1-2i) g^i} \sinh(k^i t^i)
\end{array}  \right),
\end{array} 
\end{eqnarray}
where sums and products are on $i=0,1$.

\begin{figure}
\includegraphics[scale=0.35,trim=0.3cm 0cm 3cm 1cm,clip = true]{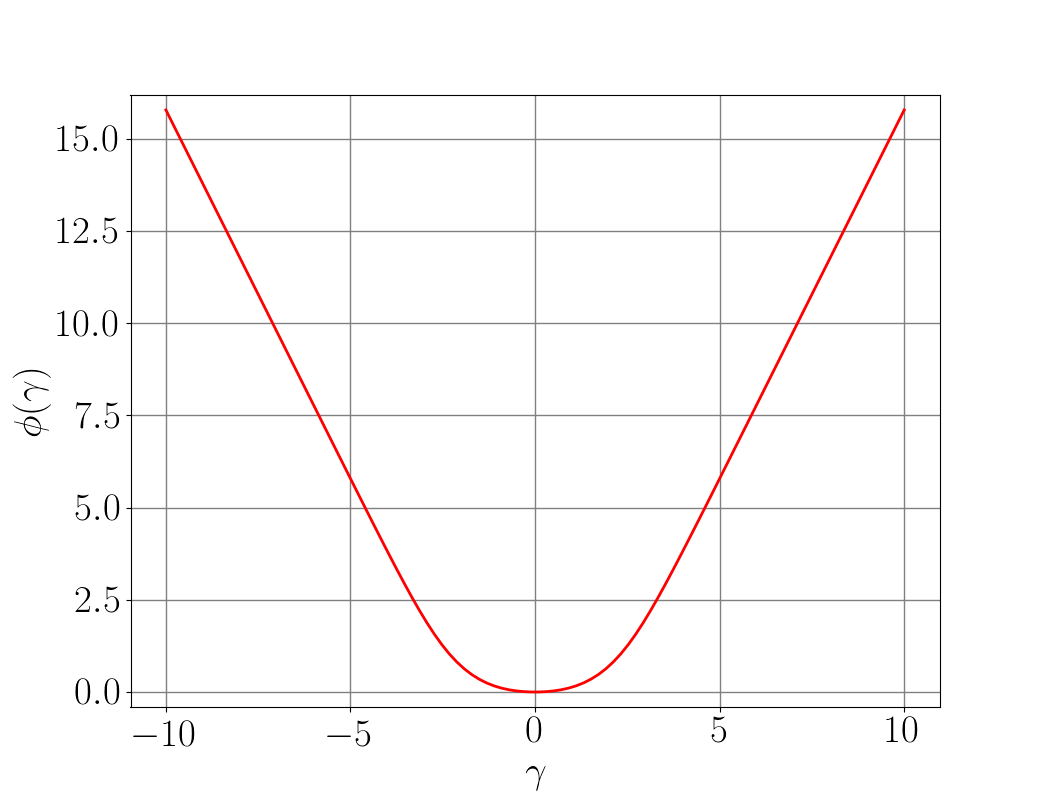}
\hfill 
\includegraphics[scale=0.35,trim=1.3cm 0cm 4cm 1cm,clip = true]{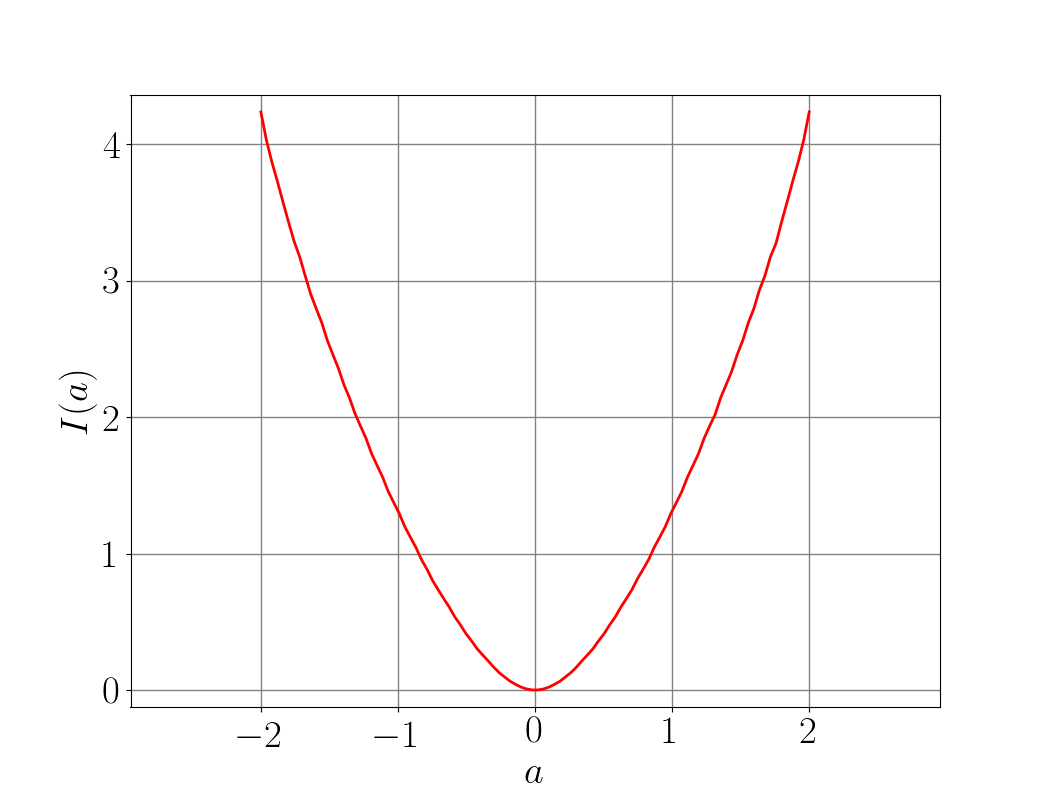} 
\caption{(left) SCGF $\phi(\gamma)$ and (right) LDF $I(a)$. 
The figures are obtained for $\alpha = 0.3$, $T = 1$, $k^0 = 1$, $k^1 = 0.1$, $g^0 = 1$, $g^1 = -1$.  \label{LDF_et_SCGF}}
\end{figure}

The highest eigenvalue of the propagator over one period writes 
\begin{equation} \label{rho_ex}
\rho_T = \frac{1}{2} \left[\mathrm{Tr} ~ \bQ_{\bkappa}(T,0) + \sqrt{\left[ \mathrm{Tr} ~ \bQ_{\bkappa}(T,0)\right]^2 - 4 \, \mathrm{Det} \bQ_{\bkappa}(T,0)} \, \right],
\end{equation}
where $\mathrm{Tr}$ and $\mathrm{Det}$ stands for the trace and determinant respectively. 
Using Eq.~\eqref{scgf_plus_grande_vp}, the SCGF $\phi(\gamma)$ follows, see Fig.~\ref{LDF_et_SCGF} for a numerical computation. The Legendre conjugate LDF $I(a)$ is shown on the same figure.
Notice that $I$ vanishes at $a = 0$ due to the symmetry of the rate matrix $\kk$: there is asymptotically as many transitions $| + \rangle \rightarrow |-\rangle $ than transitions $| - \rangle \rightarrow |+\rangle$ leading to a vanishing typical value for $A$.

\begin{figure}[b] 
\begin{center}
\includegraphics[scale=0.3,trim=3cm 0cm 4cm 2cm,clip = true]{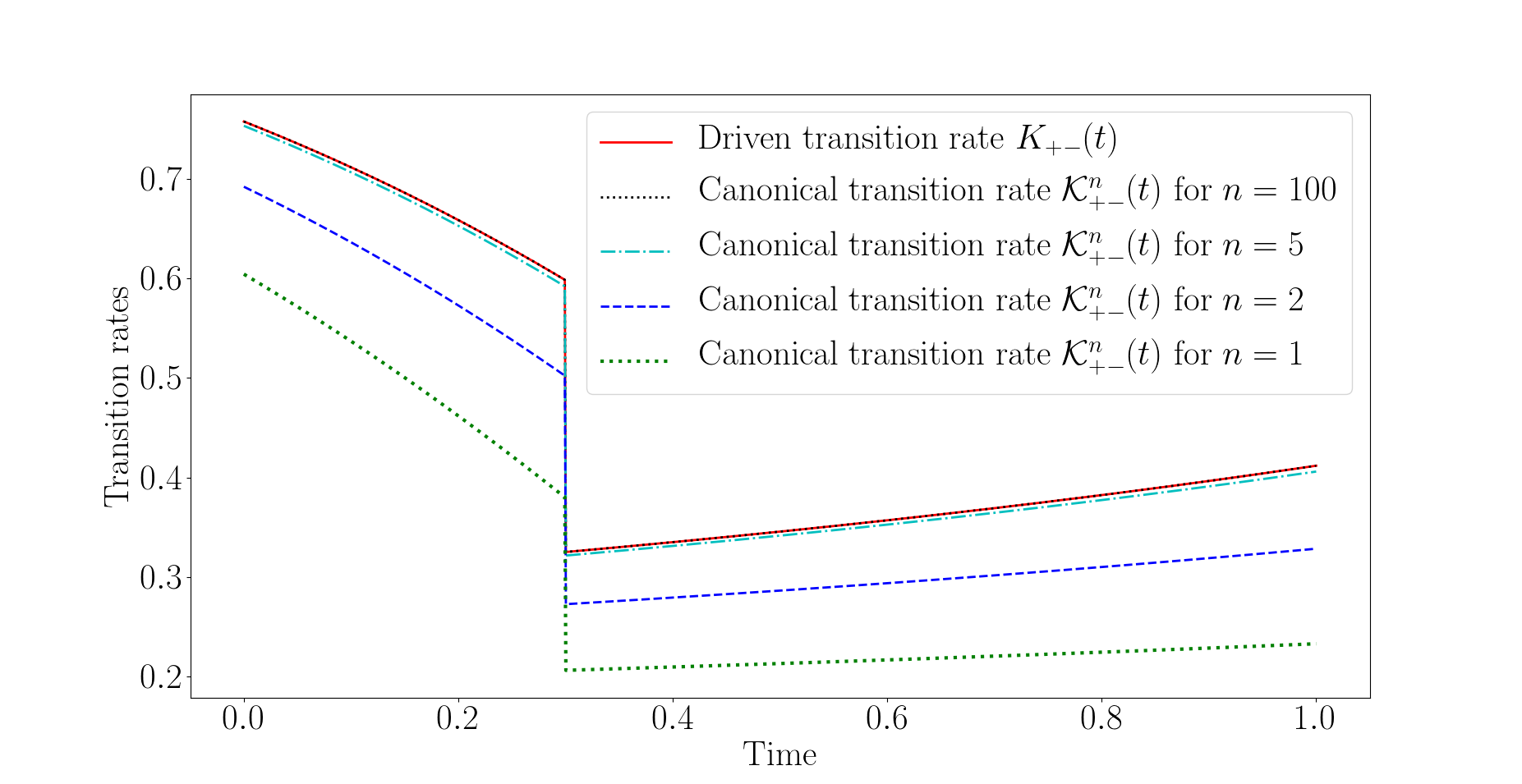}
\caption{$K_{+-}(t)$ and $\kc_{+-}(t)$ as a function of time for different number of periods $n = 1, 2, 5, 100$. The figure is obtained for $\alpha = 0.3$, $T = 1$, $k^0 = 1$, $k^1 = 0.1$, $g^0 = 1$, $g^1 = -1$, $a = 0.4$ corresponding to $\gamma = 0.85$. \label{figure_rates}}
\end{center}
\end{figure}
\begin{figure}
\begin{center}
\includegraphics[scale=0.3,trim=2cm 0cm 4cm 2cm,clip = true]{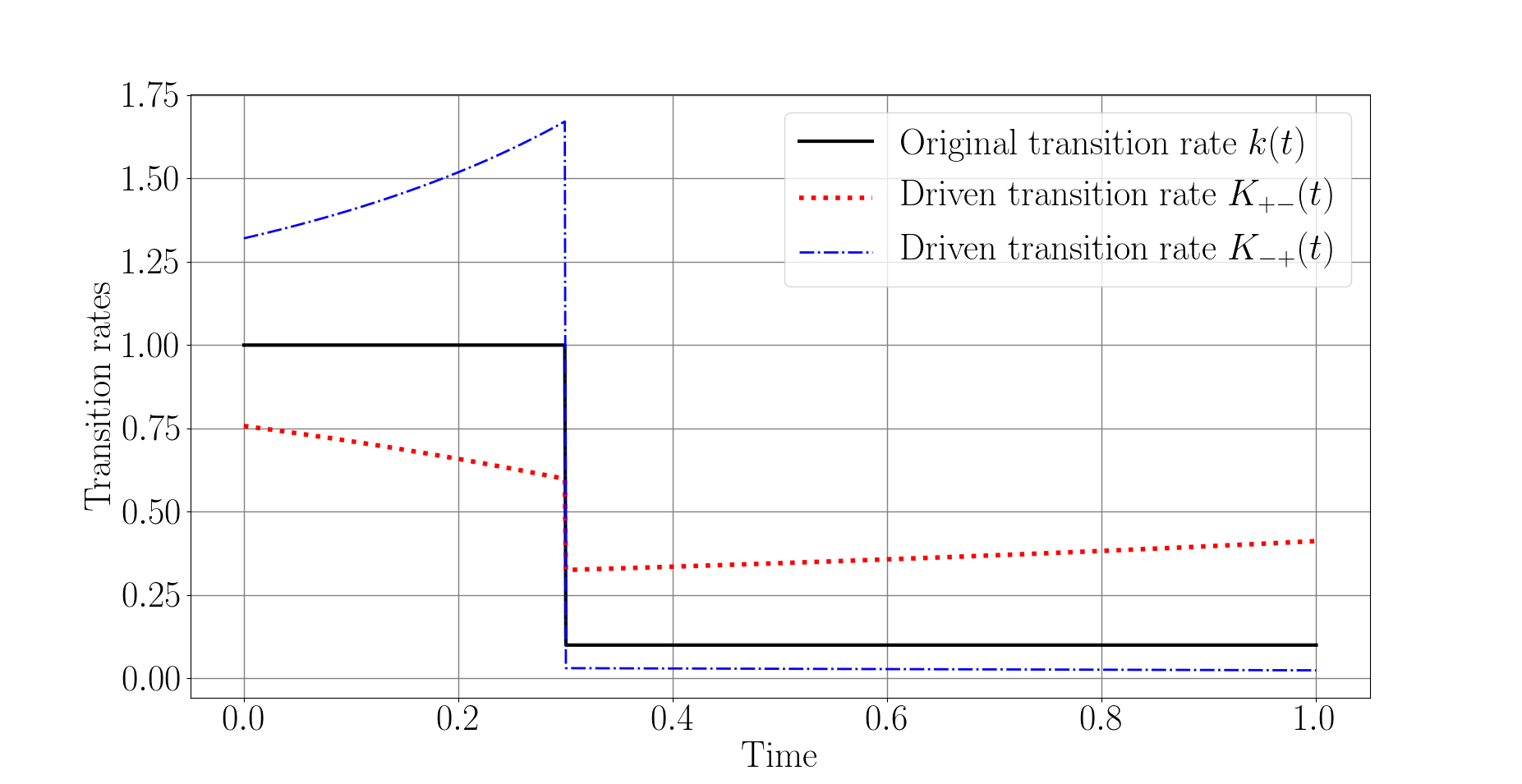}
\caption{Original transition rate $k(t)$ (solid black line) and driven transition rates $K_{+-}(t)$ (red dotted line) and $\K_{-+}(t)$ (blue dashed line). 
The figure is obtained for $\alpha = 0.3$, $T = 1$, $k^0 = 1$, $k^1 = 0.1$, $g^0 = 1$, $g^1 = -1$, $a = 0.4$ corresponding to $\gamma = 0.85$. \label{figure_rates2}}
\end{center}
\end{figure}
The generator $\K$ being defined as the Doob transform of $\bkappa$ based on $\h(t) = \h(0) [ \bQ_{\bkappa}(t,0) ]^{-1}$, we need the left eigenvector $\h(0)=\h_{T}$ of the one period propagator associated to eigenvalue $\rho_T$:
\begin{equation} \label{ex_h(0)}
\h(0) = \frac{1}{\mathcal{N}} \left( \begin{array}{c}
\prod_{i} \e^{-k^i t^i} \Big[ \sum_{i} \e^{\g g^i} \sinh(k^i t^i) \cosh(k^{1-i} t^{1-i}) \Big] \\
\rho_T - \prod_{i} \e^{-k^i t^i} \Big[ \prod_{i} \cosh(k^i t^i) + \prod_{i} \e^{\g (1-2i) g^i} \sinh(k^i t^i) \Big]
\end{array} \right)^T,
\end{equation}
with $\mathcal{N}$ a normalizing factor following from Eq.~\eqref{pdriven_normal}. Inverting the propagators in Eqs.~(\ref{Q0} -- \ref{Q1}), we can compute $\h(t)$ at any $t \in [0,\peri[$. Then, Eq.~\eqref{Doob transform_Kh} yields an analytic expression for the generator of the driven process from which we have computed numerically one component shown on Fig.~\ref{figure_rates}. 
Similarly, the generator $\kkc$ is defined as the Doob transform of $\bkappa $ based on $\C(t) = \1 ~ [\bQ_{\bkappa}(\peri,0)]^{n} [\bQ_{\bkappa}(t,0) ]^{-1}$. Inverting the propagators of Eqs.~(\ref{Q0} -- \ref{Q1}) and taking the $n$th power of the one period propagator, we can compute $\C(t)$ at any $t\in [0,\peri[$ for $n \in \mathbb{N}$. Then, Eq.~\eqref{doob_kcano} yields an expression for the generator of the driven process from which we have computed numerically one component shown on Fig.~\ref{figure_rates} for $n=1,2,5$ and $100$. This figure illustrates the convergence of the canonical generator $\kkc$ towards the driven generator $\K$ when $n \rightarrow \infty$ as stated in Eq.~\eqref{kcano_K_lim}. We observe that the two generators $\K$ and $\kkc$ are piecewise continuous (with discontinuities at phases $\alpha T$ and $T$) and time-dependent even though the original rate matrix $\kk$ was piecewise constant.

\begin{figure}
\begin{center}
\includegraphics[scale=0.3,trim=2cm 0cm 4cm 2cm,clip = true]{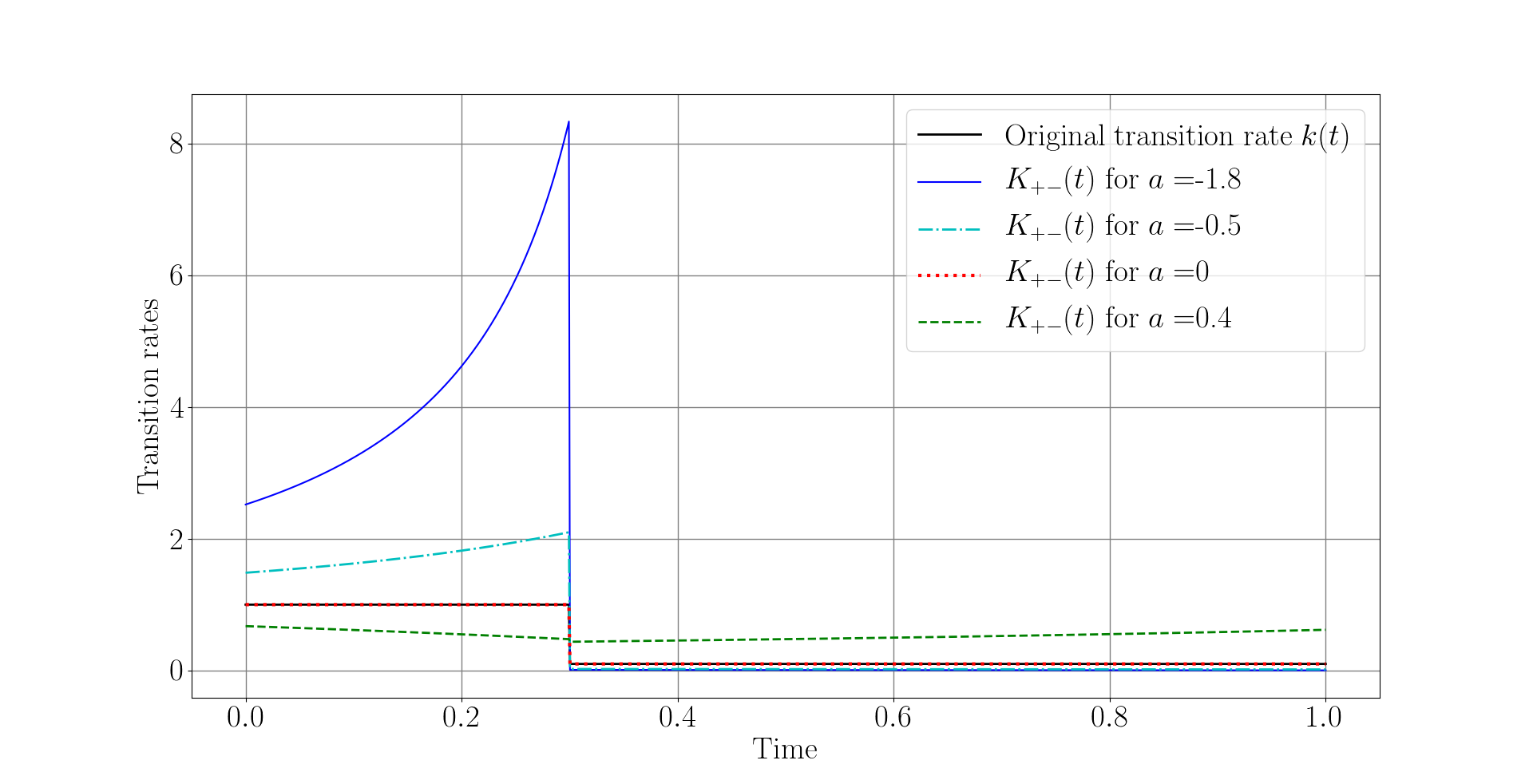}
\caption{Original transition rate $k(t)$ (solid black line) and driven rate $K_{+-}(t)$ (colored lines) for different values of $\g$ corresponding to different values of fluctuation $a$. 
The figure is obtained for $\alpha = 0.3$, $T = 1$, $k^0 = 1$, $k^1 = 0.1$, $g^0 = 1$, $g^1 = -1$. The values $a = -1.8, -0.5, 0, 0.4$ correspond respectively to $\gamma = -3.59, -1.29, 0, 1.11$ respectively. As expected, $\kk = \K$ when conditioning at the mean value $a=0$. \label{figure_rates3}}
\end{center}
\end{figure}

On Fig.~\ref{figure_rates2}, we plot both driven rates $K_{+-}(t)$ and $K_{-+}(t)$ and original rates $k_{+-}(t)=k_{-+}(t)=k(t)$ to observe qualitatively the effect of the conditioning on our initial Markov process. We chose to impose $a = 0.4$ net transitions from $|+\rangle $ to $ |-\rangle $ per unit time, counted positively if they occur on the first part of each cycle ($g^{0} = 1$) and negatively on the second part ($g^{1} = -1$). In view of $I$'s strict convexity, the process that has $a=0.4$ as a typical event is the driven process for $\gamma = 0.85 = I'(0.4)$. In the original process, $A$ is zero on average due to the symmetry of the rate matrix. Hence, imposing $a>0$ should increase the rate of the driven process for  transitions $|+\rangle \rightarrow |-\rangle$ on $[0,\alpha T]$ and transitions $|-\rangle \rightarrow |+\rangle$ on $[\alpha T, T]$. Compared to the original rate $k$, we see on Fig.~\ref{figure_rates2} that indeed $K_{+-} < k^{0} < K_{-+}$ on $[0,\alpha T]$ so that transitions $|+\rangle \rightarrow |-\rangle$ are prefered on average, and conversely $K_{-+} < k^{1} < K_{+-}$ on $[\alpha T,T]$ so that transitions $|-\rangle \rightarrow |+\rangle$ are prefered on average. Hence, the conditioning has broken the symmetry of the rate matrix and made it fully time dependent.

On Fig.~\ref{figure_rates3}, we plot the rate $K_{+-}$ for different values of $a$ (associated to their corresponding $\gamma$). We observe that this rate from $|-\rangle \rightarrow |+\rangle$ deviates more and more from $k_{+-}$ as $|a|$ becomes larger, i.e. goes away from the mean $0$ for the original process. The magnitude of change of the driven rate is thus in direct correspondence with the magnitude of the conditioning. However, it is not intuitive to understand the growth of the transition rate. We can just say that the possibility of a time-dependent rate matrix offers a broader dynamical space to explore in the variational calculation compared to case of piecewise-constant rates.

\section{Conclusion}

Beyond the computation of the cumulants of a random variable $\A$, recent developments in large deviation theory provide a mathematical framework to study path probabilities conditioned on an event $\A=\bm{a}$ (microcanonical conditioning, $\A$ does not fluctuate). One aims to build a new Markov process for which $\A$ converges in probability to the value $\ba$ at long time (canonical conditioning, $\A$ fluctuates). In this paper, we addressed this problem of process conditioning for observables defined through periodic functions in the framework of Markov jump processes with time-periodic generators. We took the period of these functions equal to the period of the generator, with no loss of generality compared to the case of commensurable periods. We focused on jump processes, but we expect our results to be transposable to general Markov processes.
Starting from nonequilibrium path probabilities generalizing the canonical and microcanonical ensembles, we defined the Markov generator of the canonical process and its asymptotic equivalent after a large number of periods. The latter is the driven generator obtained from the Doob transform involving an eigenvector of the one-period propagator for the tilted operator (and its time evolution). This is consistent with the stationary theory where an eigenvector of the tilted matrix is involved instead. Finally, the conditioned-free process for which $\A$ takes asymptotically the microcanonical value $\ba$ follows from the driven process.
This result requires the ensemble equivalence between microcanonical and canonical path ensembles which is granted by the convexity of the LDF for $\A$, in straight connection with entropy's concavity for the equivalence of equilibrium ensembles. This analogy between entropy and LDF is broader than the question of ensemble equivalence. In the same way that the canonical state probability follows from Jayne's Maximum entropy principle in equilibrium statistical mechanics, the driven process follows from a constrained optimization problem on the 2.5 LDF.

\section*{Acknowledgement}

We thank H. Vroylandt for his pertinent comments on the definition of the Doob transform via the variational approach.

\appendix
\section*{Appendix}
\addcontentsline{toc}{section}{Appendices}

\section{Definition of the Doob transform} \label{Doob}

A linear system of first order differential equations does not conserve norm in general. Let $\M$ be an arbitrary Metzler matrix and $\bv$ a vector whose elements are strictly positive. The Doob transform of $\M$ associated with $\bv$ is defined by
\begin{equation} \label{DoobTransform}
\M^{\bv} \doteq \mathcal{D}(\bv) \M \, \mathcal{D}(\bv)^{-1} - \mathcal{D}(\bv)^{-1} \mathcal{D}(\bv \M),
\end{equation}
where $\mathcal{D}(\bv)$ is the diagonal matrix with the components of $\bv$ on its diagonal. Componentwise, Eq.~\eqref{DoobTransform} writes 
\begin{equation}
M^{\bv}_{xy}(t) = v_x(t) M_{xy}(t) v_y^{-1}(t) - v_x^{-1}(t) (\bv \M)_x(t) \delta_{xy}.
\end{equation}
$\M^{\bv}$ generates an honest process since $\sum_x M^{\bv}_{xy} = 0$. The Doob transform is then a tool to build an honest process out of a non-honest one. Notice that if $\alpha$ is a constant, $\M^{\alpha \bv} = \M^{\bv}$. This definition of the Doob transform is a special case of a generalized definition~\cite{Chetrite2014}. 
The path probability associated with the Doob transform $\M^{\bv}$ is given by~\cite{Chetrite2011}
\begin{equation} \label{path_proba_doob}
\mathds{P}_{\M^{\bv},\bpi(0)}[z] = \mathds{P}_{\M,\bpi(0)}[z] \, v_{x(\tf)}(\tf) \exp\left[ - \int_0^{\tf} \left(v_{x(t)}^{-1}(t)(\bv \M)_{x(t)}(t) + v_{x(t)}^{-1}\left.  \frac{\partial \bv(t)}{\partial t}\right|_{x(t)} \right) \ud t \right] v_{x_0}^{-1}(0),
\end{equation}
where the second term in the integrand is due to the time dependence of vector $\bv$ leading to $\ln v_{x_{i}}(t_{i+1}) - \ln v_{x_{i}}(t_{i}) = \int_{t_{i}}^{t_{i+1}} dt \partial_{t} \ln v_{x_{i}}(t)$ contributions for each interval of time $[t_{i},t_{i+1}]$ between two jumps.

\section{SCGF from the optimizer} \label{app_scgf}

We recover the SCGF for observable $A$ by evaluating the 2.5 LDF at the optimum $(\om,\p)$ of our variational problem stated at Eqs.~(\ref{I(a)}). By definition of the second component of our observable $\A$ in Eq.~\eqref{observabletest} and using Eq.~\eqref{opt_p} we find
\begin{eqnarray}
\gamma_2 ~ A_2(\p) & = & \frac{1}{\peri} \int_0^\peri \sum_{x} p_x ~ \gamma_2 \, f_x , \\
& = & \frac{1}{\peri} \int_0^\peri \left[ c + \sum_{x} p_x \left[\lambda_x - \Lambda_x \right] -  \sum_{x} p_x \dot{u}_x \right], \\
& = & \frac{1}{\peri} \int_0^\peri \left[ c + \sum_{x} p_x \left[\lambda_x - \Lambda_x \right] + \sum_{x} \dot{p}_x u_x \right], \label{j2(opt)}
\end{eqnarray}
where we used \textbf{C3} in the integration by part. Using Eqs.~(\ref{contraction_om2}, \ref{K}) and \textbf{C2}, the LDF at the optimum writes
\begin{eqnarray}
I(\om,\p) & = & \frac{1}{\peri} \int_0^\peri \left[\sum_{x} p_x \left[\lambda_x - \Lambda_x \right] + \sum_{x, y \neq x} \omega_{xy} \left[u_x - u_y\right] + \gamma_1 \sum_{x, y \neq x} \omega_{xy} \, g_{xy} \right],  \\ 
& = & \frac{1}{\peri} \int_0^\peri \left[ \sum_{x} p_x \left[\lambda_x - \Lambda_x \right]  + \sum_{x, y \neq x} u_{x} \left[\omega_{xy} - \omega_{yx}\right] + \gamma_1 \sum_{x, y \neq y} \omega_{xy} \, g_{xy}\right], \\
& = & \frac{1}{\peri} \int_0^\peri \left[ \sum_{x} p_x \left[\lambda_x - \Lambda_x \right] + \sum_{x} u_{x} \, \dot{p}_x \right] + \gamma_1 A_1 \label{I(opt)}.
\end{eqnarray}
Combining Eqs.~\eqref{j2(opt)} and \eqref{I(opt)}, we finally obtain
\begin{eqnarray}
\gamma_1 ~ A_1(\K \circ \p) + \gamma_2 ~ A_2(\p)  - I_{2.5}(\K \circ \p, \p)  =  \frac{1}{\peri} \int_0^\peri c(\tau) \ud \tau = \phi(\g).
\end{eqnarray}
The left-hand-side is the SCGF as the Legendre transform of the LDF. It follows that the SCGF is the time-average over a period of the Lagrange multiplier used to normalize the occupation density, recovering the result of Eq.~\eqref{expu_vp}. As mentioned in the conclusion, the variational calculation of the SCGF is similar in many ways to the calculation of equilibrium canonical probability via the maximum entropy principle in which the SCGF (free energy) is also connected to the Lagrange multiplier that imposes probability normalization.

\section{Definitions of the time-ordered exponential} \label{app_propagator}

The ordered exponential $\bQ_{\M}(t,0) \doteq \overleftarrow{\exp} \int_{0}^t \M(t') \, \ud t'$ is the unique solution of the initial value problem:
\begin{equation} \label{timeorderedexp}
\frac{\ud}{\ud t}\X (t) = \M(t) \X(t), \quad \text{with }
\X(0) = \mathbb{1},
\end{equation}
that has the integral form 
\begin{equation} \label{timeorderedexpint}
\X(t) = \mathbb{1} + \int_0^t \M(t') \X(t') \ud t'.
\end{equation}
Using this integral form into itself, one obtains the series expansion of the time ordered exponential
\begin{eqnarray}
\bQ_{\M}(t,0) & = & \mathbb{1} + \int_0^t \M(t_1) \ud t_1 + \int_0^t \ud t_1 \int_0^{t_1}  \ud t_2 \, \M(t_1)  \M(t_2) \nonumber \\
& & +  \int_0^t \ud t_1 \int_0^{t_1}  \ud t_2 \int_0^{t_2}  \ud t_3 \, \M(t_1) \M(t_2) \M(t_3) \, + \, \ldots
\end{eqnarray}
Notice that the arrow on the exponential specifies the ordering of the product of $\M$ in the expansion for increasing time from right to left. 

The reverse-ordered exponential $\boldsymbol{\overrightarrow{\mathcal{Q}}}_{\M}(0,t) \doteq \overrightarrow{\exp} \int_{0}^t \M(t') \, \ud t'$ is unique solution of the initial value problem:
\begin{equation} \label{reversetimeorderedexp}
\frac{\ud}{\ud t} \X(t) = \X(t) \M(t), \quad \text{with } \X(0) = \mathbb{1},
\end{equation}
that has the integral form
\begin{equation} \label{reversetimeorderedexpint}
\X(t) = \mathbb{1} + \int_0^t \X(t') \M(t') \ud t'.
\end{equation}
Using this integral form into itself, one obtains the series expansion of the reverse-ordered exponential
\begin{eqnarray}
\boldsymbol{\overrightarrow{\mathcal{Q}}}_{\M}(0,t) & = & \mathbb{1} + \int_0^t \M(t_1) \ud t_1 + \int_0^t \ud t_1 \int_0^{t_1}  \ud t_2 \, \M(t_2) \M(t_1) \nonumber \\
& & + \int_0^t \ud t_1 \int_0^{t_1}  \ud t_2 \int_0^{t_2}  \ud t_3 \, \M(t_3) \M(t_2) \M(t_1) \, + \, \ldots
\end{eqnarray}
Notice that the arrow on the exponential specifies the ordering of the product of $\M$ in the expansion for increasing time from left to right. 

\section{Properties of the time-ordered exponential} 
\label{Properties}

For the reader convenience, we remind useful properties on linear differential equations with periodic generators. See Ref.~\cite{Adrianova1995} for a full description of the theory.

\begin{property}[Transpose of a propagator] \label{P1} The transpose of a propagator based on generator $\M$ is the time reverse propagator based on the transposed generator $\M^{T}$
	\begin{equation}
		\left[\bQ_{\M}(t,0) \right]^T = \boldsymbol{\overrightarrow{\mathcal{Q}}}_{\M^T}(0,t).
	\end{equation}
\end{property}
The property follows from the definitions and the fact that the transpose of a product of two matrices is the reverse product of the two transposed matrices.

\begin{property}[Inverse of a propagator] \label{P2} The inverse of a propagator based on generator $\M$ is the time reverse propagator based on the opposite generator $-\M$
\begin{equation}
\Big[\bQ_{\M}(t,0) \Big]^{-1} = \boldsymbol{\overrightarrow{\mathcal{Q}}}_{-\M}(0,t).
\end{equation}
\end{property}
Assuming $\dot{\X} = \M \X $ and since $\frac{\ud}{\ud t} (\X\X^{-1}) = \dot{\X}\X^{-1} + \X\dot{\X}^{-1} = 0$, we have $\dot{\X}^{-1} = - \X^{-1} \dot{\X} \X^{-1} = - \X^{-1} \M $. Hence the two propagators are connected.

\begin{property}[First relation between final and initial value problems] \label{P3}
The solution of the final value problem 
\begin{equation} 
\frac{\ud}{\ud t}\X (t) = - \X(t) \M(t), \quad \text{with }
\X(T) = \mathbb{1},
\end{equation}
is given by $\X(t) = \bQ_{\M}(T,t)$. 
\end{property}
Indeed, one can check directly that
\begin{eqnarray}
\frac{\ud}{\ud t} \bQ_{\M}(T,t) 
& = & \lim_{a \rightarrow 0} \frac{\bQ_{\M}(T,t+a)-\bQ_{\M}(T,t+a) \, \bQ_{\M}(t+a,t)}{a}, \\ 
& = & \lim_{a \rightarrow 0} \bQ_{\M}(T,t+a) \times \frac{\mathbb{1} - \bQ_{\M}(t+a,t)}{a}, \\
& = & \bQ_{\M}(T,t) \times \Big[ - \frac{\ud}{\ud s} \bQ_{\M}(s,t) \mid_{s = t} \Big], \\
& = & - \bQ_{\M}(T,t) ~ \M \bQ_{\M}(t,t), \\
& = & - \bQ_{\M}(T,t) ~ \M.
\end{eqnarray}

\begin{property}[Second relation between final and initial value problems] \label{P4}
The solution of the final value problem:
\begin{equation} 
\frac{\ud}{\ud t}\X (t) = \M(t) \X(t), \quad \text{with }
\X(T) = \mathbb{1},
\end{equation}
is given by $\X(t) = \Big[ \bQ_{\M}(T,t) \Big]^{-1} = \boldsymbol{\overrightarrow{\mathcal{Q}}}_{-\M}(t,T)$.
\end{property}
This follows from combining properties \ref{P2} and \ref{P3}.

\begin{property}[Time-ordered exponential of the sum of commuting matrices] \label{P5}
If $\M(t_1)$ and $\N(t_2)$ commute for any $t_1$,$t_2 \in \R$, then $ \bQ_{\M + \N}(t,t_0) = \bQ_{\M}(t,t_0) \bQ_{\N}(t,t_0)$.
\end{property}
Let us denote the left-hand side of the equality by $\X(t)$ and the right-hand side by $\Y(t)$. On the one hand, $\dot{\X} = (\M + \N) \X$. On the other hand, $\dot{\Y} = \M \Y + \bQ_{\M}(t,t_0) \N \bQ_{\N}(t,t_0) = (\M+\N) \Y$ since $\M$ and $\N$ commute for any time. Thus, the matrices $\X$ and $\Y$ satisfy the same matrix differential equation. Besides, $\X(t_0) = \Y(t_0) = \mathbb{1}$, hence $\X(t) = \Y(t)$, $\forall t \in \R$.

\bibliography{BibDriven}

\end{document}